\documentclass[aps,prl,reprint,10pt,twocolumn]{revtex4-1}
\usepackage{graphicx}
\usepackage{amssymb}
\usepackage{amsmath}
\usepackage{amsfonts}
\usepackage{color}
\usepackage[latin1]{inputenc}
\usepackage{float}
\usepackage{comment}

\begin{document}
\title{Kardar-Parisi-Zhang growth on one-dimensional decreasing substrates}

\author{I. S. S. Carrasco}
\email{ismael.carrasco@ufv.br}
\author{T. J. Oliveira}
\email{tiago@ufv.br}
\affiliation{Departamento de F\'isica, Universidade Federal de Vi\c cosa,
36570-900, Vi\c cosa, Minas Gerais, Brazil}

\begin{abstract}
Recent experimental works on one-dimensional (1D) circular Kardar-Parisi-Zhang (KPZ) systems whose radii \textit{decrease} in time have reported controversial conclusions about the statistics of their interfaces. Motivated by this, we investigate here several 1D KPZ models on substrates whose size changes in time as $L(t)=L_0 + \omega t$, focusing on the case $\omega<0$. From extensive numerical simulations, we show that for $L_0 \gg 1$ there exists a transient regime in which the statistics is consistent with that of flat KPZ systems (the $\omega=0$ case), for both $\omega<0$ and $\omega>0$. Actually, for a given model, $L_0$ and $|\omega|$, we observe that a difference between ingrowing ($\omega<0$) and outgrowing ($\omega>0$) systems arises only at long times ($t \gtrsim t_c=L_0/|\omega|$), when the expanding surfaces cross over to the statistics of curved KPZ systems, whereas the shrinking ones become completely correlated. A generalization of the Family-Vicsek scaling for the roughness of ingrowing interfaces is presented. Our results demonstrate that a transient flat statistics is a general feature of systems starting with large initial sizes, regardless their curvature. This is consistent with their recent observation in ingrowing turbulent liquid crystal interfaces, but it is in contrast with the apparent observation of curved statistics in colloidal deposition at the edge of evaporating drops. A possible explanation for this last result, as a consequence of the very small number of monolayers analyzed in this experiment, is given. This is illustrated in a competitive growth model presenting a few-monolayer transient and an asymptotic behavior consistent, respectively, with the curved and flat statistics.
\end{abstract}


\maketitle


Some universality classes for nonequilibrium interface growth are known to split into subclasses depending on the initial conditions (ICs) of the growth. More specifically, given an evolving interface, the height $h(\vec{x},t)$ at a given position $\vec{x}$ and time $t$ is expected to fluctuate according to universal height distributions (HDs). While the variance of the HDs - the squared interface width $w_2$ - and the correlation length parallel to the substrate $\xi$ increase in time, respectively, as $w_2 \sim t^{2\beta}$ and $\xi \sim t^{1/z}$, with the universal growth ($\beta$) and dynamic ($z$) exponents defining the universality class in a given dimension, the HDs' probability density functions are dependent on the ICs. This interesting feature, firstly demonstrated by Prah\"ofer and Sphon \cite{Prahofer2000} in the solution of the one-dimensional (1D) polynuclear growth (PNG) model, which belongs to the Kardar-Parisi-Zhang (KPZ) class \cite{kpz}, has been numerically observed also in 2D KPZ class \cite{healy12,*tiago13,*healy13}, as well as in another class relevant for thin film deposition by molecular beam epitaxy (MBE) \cite{Ismael16a}.

In the solution of the 1D PNG model \cite{Prahofer2000}, beyond the natural stationary (\textit{Brownian}) IC, there are other two relevant ICs, both of which leads to HDs given by Tracy-Widom (TW) \cite{TW} distributions from random matrix theory. For {flat} IC, the growth starts on an initially flat substrate of large size $L_0$, which does not change in time, and the HD is given by the Gaussian orthogonal ensemble (GOE) TW distribution. In the {droplet} IC, the substrate is initially small ($L_0 \rightarrow 0$), but expands linearly in time [$L(t)=\omega t$]. In this case, the Gaussian unitary ensemble (GUE) TW distribution sets the asymptotic KPZ HD. Since the macroscopic shape of the PNG interfaces is curved for the droplet IC, it was conjectured that the splitting of the KPZ class is related with the surface geometry. Indeed, the universality of GOE and GUE HDs have been widely confirmed theoretically \cite{SasaSpo1,*Amir,*Calabrese2011}, experimentally \cite{Takeuchi2010,*Takeuchi2011} and numerically \cite{Alves11,*tiago12a,*Alves13,healy141d,silvia17} in 1D KPZ systems with flat and curved macroscopic shapes, respectively. For additional information, see the recent reviews \cite{Kriecherbauer2010,*Corwin-RMTA2012,*HHTake2015}.

Beyond the (1-point) HDs, (2-point) spatial and temporal covariances are also universal and dependent on the ICs (or geometry). For instance, for the 1D KPZ class the spatial covariances are associated with the so-called Airy$_1$ and Airy$_2$ processes for flat and curved geometries, respectively \cite{Prahofer2002,*Sasa2005,*Borodin}. Geometry-dependent covariances have also been numerically found in the 2D KPZ class \cite{Ismael14}, as well as for the nonlinear MBE class \cite{Villain,*LDS} in both 1D and 2D \cite{Ismael16a}.

In a recent work, Fukai and Takeuchi (FT) \cite{Fukai2017} reported experimental and numerical results demonstrating that the statistics of circular KPZ interfaces with inward growth, i.e., whose average radii decrease in time, is given by GOE HDs and Airy$_1$ covariance, rather than the GUE/Airy$_2$ expected for curved interfaces. This finding is in contrast with the apparent observation of GUE fluctuations by Yunker {\it et al.} \cite{Yunker2013a} in the inward growth of anisotropic colloidal particles deposited at the edge of evaporating drops. In this Rapid Communication, we demonstrate through extensive simulations of several KPZ models on 1D size-changing substrates that ingrowing interfaces indeed present a transient flat statistics, in agreement with FT \cite{Fukai2017}, whenever their initial size $L_0$ is very large. More important, our results let clear that this transient is not a consequence of the inward growth, as Ref. \cite{Fukai2017} suggests, and exists even for outgrowing interfaces. A possible explanation for the apparent curved statistics reported in Ref. \cite{Yunker2013a} is also presented.

We start investigating three discrete models belonging to the KPZ class in 1D: the Etching model by Mello {\it et. al} \cite{Mello01}, the restricted solid-on-solid (RSOS) by Kim and Kosterlitz \cite{kk} and the single step (SS) model \cite{barabasi}. In all cases, periodic boundary conditions are considered and particles are sequentially deposited on a flat substrate, whose average size changes in time as $\langle L(t)\rangle=L_0+\omega t$, at randomly chosen positions (say, $i$). In the Etching model, $h_i\rightarrow h_i+1$ and the heights of the nearest neighbor (NN) sites are individually updated to $h_i-1$ if they are smaller than this value. In the SS model, depositions are accepted wherever $h_i$ is a local minimum and, then, $h_i \rightarrow h_i+2$. Finally, in the RSOS model, the deposition of a particle is accepted ($h_i\rightarrow h_i+1$) only if it does not yield a step $|h_i-h_{i\pm 1}| > 1$ in the interface. The growth starts with $h_i=0$ $\forall$ $i \in [1,L_0]$ for the Etching and RSOS models, while in the SS model one makes $h_i=1$ if $i$ is odd and $h_i=0$ otherwise. Following the method introduced by us in Ref. \cite{Ismael14}, the enlargement of the active growing zone (the case $\omega>0$) is implemented by simply duplicating columns at rate $\omega$. Two probabilities are defined so that in one time unity the average number of duplications is equal to $\omega$, and the average number of depositions is equal to $\langle L \rangle$ in that interval. Namely, at each time step $\Delta t=1/(L+\omega)$, one particle is deposited with probability $P_d=L/(L+\omega)$ or a column is duplicate with complementary probability $P_{\omega}=\omega/(L+\omega)$. The duplications are implemented by randomly choosing a column $i$ and then creating a copy of it in the {\it new} position $i+1$. Since the height differences in the SS model are restricted to $|h_i-h_{i-1}|=1$, we have to duplicate a pair of NN columns and, then, $\omega/2$ is used in the definition of the probabilities. To investigate systems with inward growth (i.e., the case $\omega<0$, which leads to decreasing substrate sizes) instead of duplicating columns, one removes (randomly chosen) columns at rate $|\omega|$. Once again, in the SS model a pair of columns has to be removed. Moreover, only pairs of columns that, after the remotion, keep the height differences $|h_i-h_{i-1}|=1$ are chosen. Similarly, in the RSOS model, only removings that do not yield steps larger than one are made.

\begin{figure}[t]
\includegraphics[width=8.5cm]{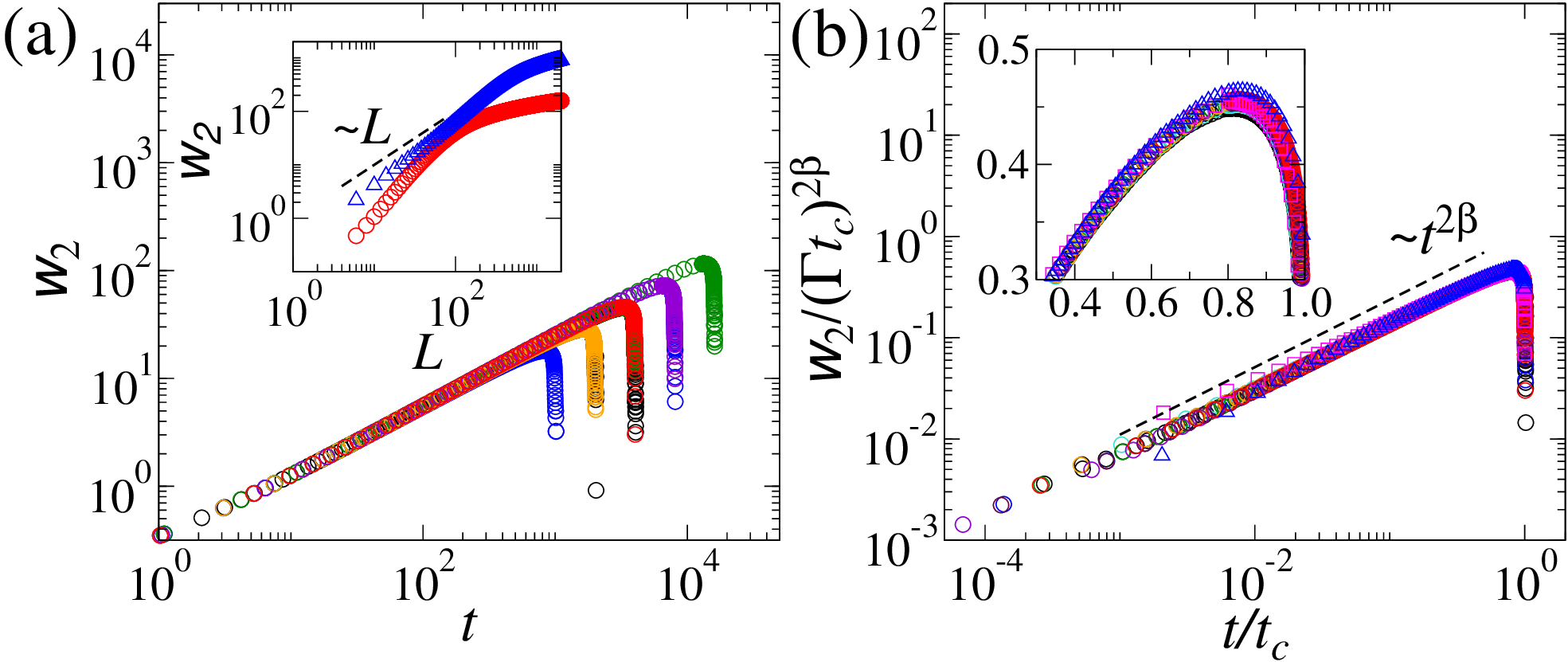}
\includegraphics[width=8.5cm]{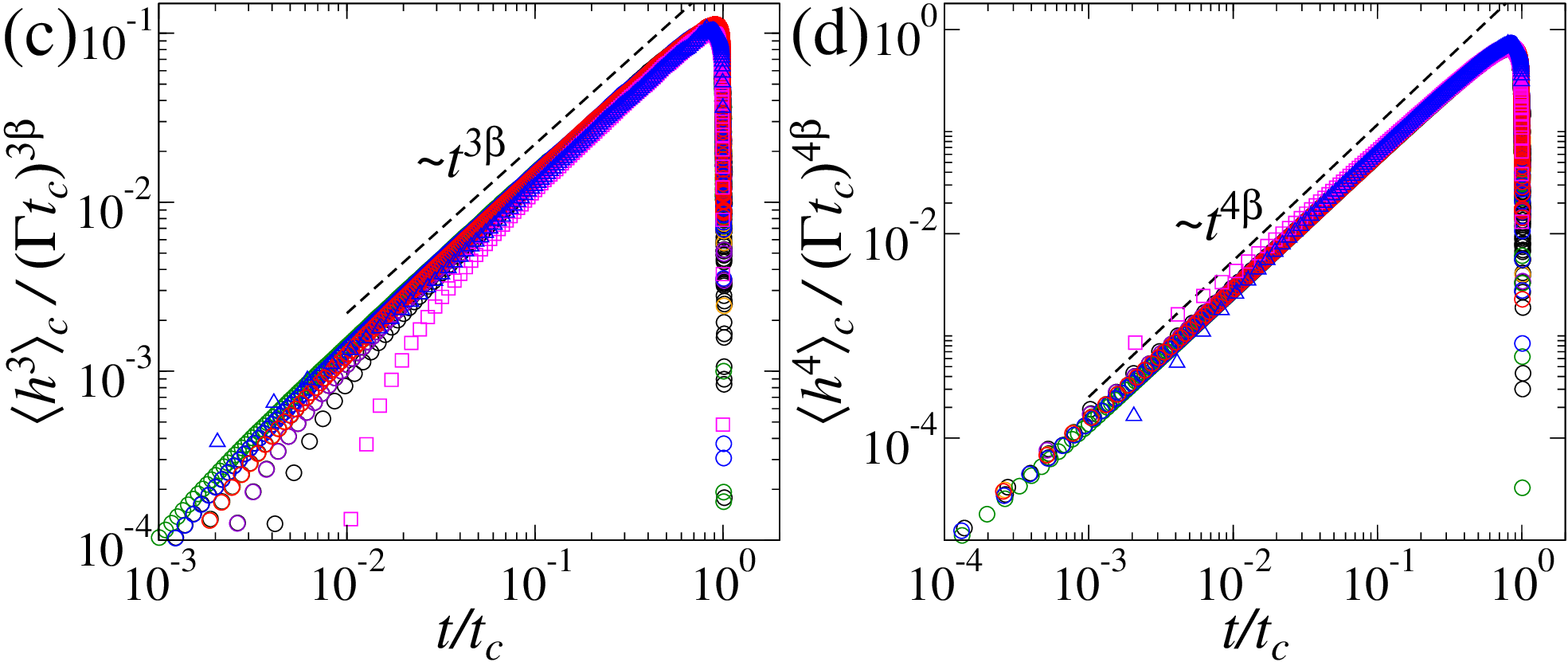}
\includegraphics[width=8.5cm]{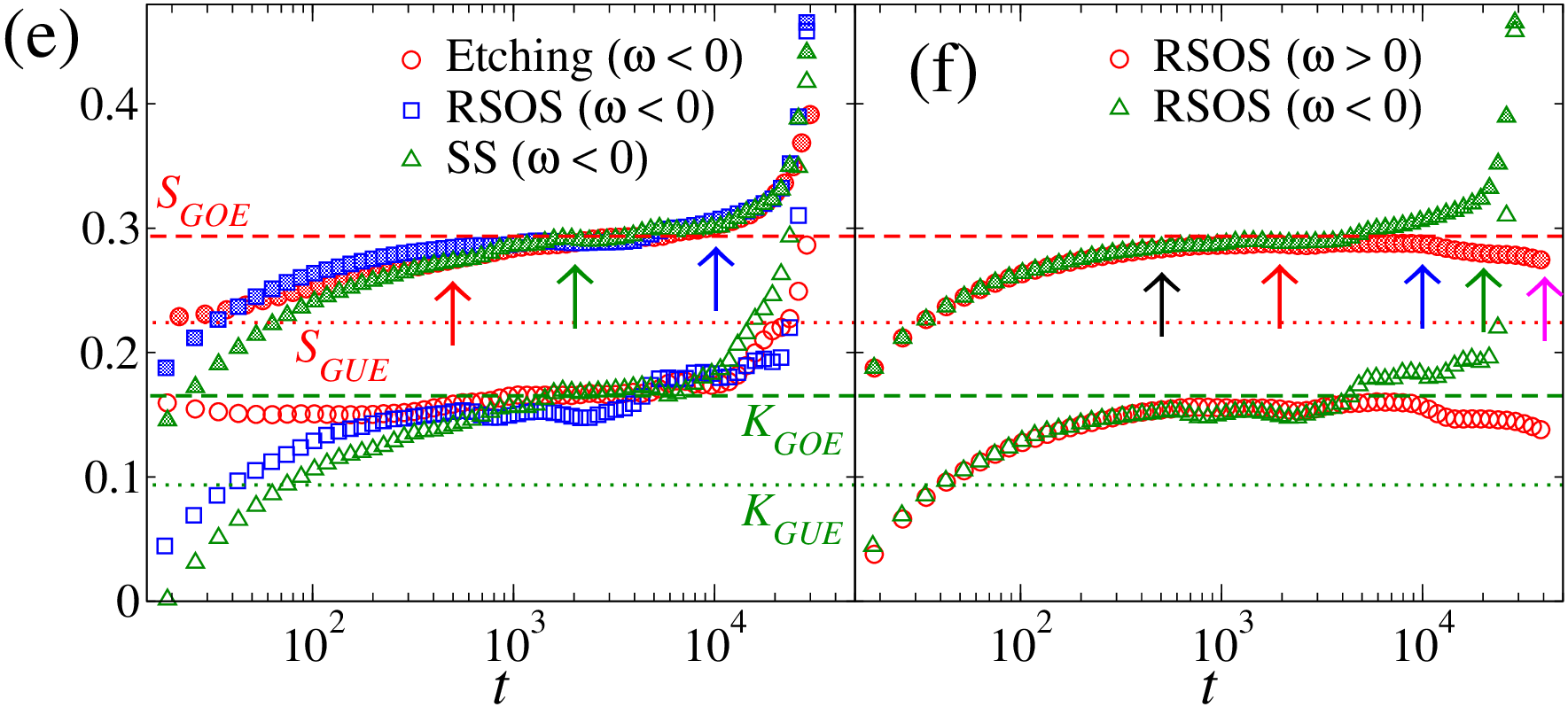}
\caption{(Color online) (a) Squared interface width $w_2=\langle h^2\rangle_c$ versus time for the RSOS model, with $\omega=-10$, $-20$ and $-40$ and $L_0=4\times$, $8\times$ and $16\times10^4$. Rescaled HDs' cumulants $\langle h^n\rangle_c/(\Gamma t_c)^{n\beta}$ against rescaled time $t/t_c$ for (b) $n=2$, (c) $n=3$ and (d) $n=4$, for the RSOS (circles), Etching (triangles) and SS (squares) models, and several values of $\omega<0$ and $L_0$.
The dashed lines have the indicate slopes, with $\beta=1/3$. The insertion in (a) shows $w_2$ versus $L$, when $t \rightarrow t_c$, for the Etching and RSOS models, with $\omega=-2$ and $L_0=8\times10^4$. In (b), the same data from the main panel are depicted in the inset in linear scale, highlighting the crossover region. The temporal evolutions of the skewness $S$ and kurtosis $K$ of the HDs are shown in (e) and (f) for the indicated models, with $|\omega|=20$ and $L_0 = 6 \times 10^5$. The arrows in (e)-(f) indicate the times when the covariances shown in Fig. \ref{fig2}a-b were measured.}
\label{fig1}
\end{figure}

For each model and set of parameters ($L_0$ and $\omega$), we have calculated the HDs [$P(h,t)$] and analyzed their $n^{th}$ cumulant [$\langle h^n\rangle_c$] up $n=4$, as well as adimensional ratios of them: $R=\sqrt{\langle h^2 \rangle_c}/\langle h \rangle$ (the variation coefficient), $S=\langle h^3\rangle_c/\langle h^2\rangle_c^{2/3}$ (the skewness) and $K=\langle h^4\rangle_c/\langle h^2\rangle_c^2$ (the kurtosis). In the growth regime, when $\xi \ll L(t)$, the height at a giving point of a fluctuating interface is expected to evolve as \cite{Krug1992,Prahofer2000}
\begin{equation}
h=v_{\infty}t+s_{\lambda}(\Gamma t)^ {\beta} \chi + \dots,
\label{eqanzatz}
\end{equation}
where $v_{\infty}$, $s_{\lambda}(=\pm1)$ and $\Gamma$ are model dependent parameters, whose values for the three models analyzed here can be found, e.g., in Ref. \cite{Ismael14}. $\chi$ is a random variable given by GOE (GUE) TW distribution for flat (curved) 1D KPZ interfaces. Therefore, the cumulants of order $n\geqslant2$ should scale as $\langle h^n\rangle_c \simeq s_{\lambda}^n (\Gamma t)^{n\beta} \langle \chi^n\rangle_c$ at not so long times. This behavior, which is well-known for static and expanding substrates, is confirmed in Figs. \ref{fig1}a-d for $n=2$, $3$ and $4$ for the ingrowing case.

At a crossover time $t^*$, however, all cumulants start decreasing in time, when $\omega<0$, with the variance $\langle h^2\rangle_c$ decreasing faster than the higher order cumulants, giving rise to a fast increase in the ratios $S$ and $K$, as observed in Figs. \ref{fig1}e-f, and also in Ref. \cite{Fukai2017}. We notice that our ingrowing systems disappear [$\langle L(t)\rangle=0$] at a characteristic time $t_c=L_0/|\omega|$, which turns out to be the natural temporal scale here. In fact, rescaling the time by $t_c$ and the $n^{th}$ cumulant by $(\Gamma t_c)^{n\beta}$, a striking data collapse is found for all models and parameters, as show Figs. \ref{fig1}b-d. Hence, the crossover time $t^* \approx 0.8 t_c$ (see the insertion in Fig. \ref{fig1}b) depends only on the parameters $L_0$ and $\omega$, being independent of the model. As pointed in Ref. \cite{Fukai2017}, the long time behavior is a simple consequence of the correlation length $\xi(t)$ becoming of the same order of $\langle L(t)\rangle=L_0-|\omega|t$ in ingrowing systems. Such condition [$\xi(t^*) \approx \langle L(t^*)\rangle$] immediately leads to $t^* \approx t_c [1 - \xi(t^*)/L_0]$. Since it is not clear for us how to calculate the exact ratio $\xi(t^*)/L_0$ for ingrowing systems, to obtain an approximation for $t^*$, we will use the ratio for fixed size substrates. Following the calculations in Ref. \cite{KrugAdv} (for $\omega=0$), the crossover to the correlated state happens when $\xi(t^*)/L_0 = (12 c_2)^{-1}$, where $c_2$ is an universal constant whose value is associated with the variance of $\chi$ as $c_2 = \langle \chi^2 \rangle_c/2^{2/3}$. Assuming that $\chi$ fluctuates according to the GOE TW distribution, as confirmed in Figs. \ref{fig1}e-f, for which $\langle \chi^2 \rangle_c = 0.63805$, one finds $c_2 = 0.40195$, in agreement with the numerical estimate reported in \cite{KrugAdv}. Thence, $\xi(t^*)/L_0 = 0.20733$, and finally $t^* \approx 0.8 t_c$, which agrees quite well with the numerical result (see Fig. \ref{fig1}b).

For fixed-size substrates, in the saturation regime the squared interface width scale as $w_2 \equiv \langle h^2\rangle_c \sim L^{2 \alpha}$, with $\alpha=1/2$ being the roughness exponent for the 1D KPZ class \cite{barabasi}. Thereby, since $\langle L(t)\rangle=L_0 [1-t/t_c]$ for ingrowing systems, one could expect that $w_2 \sim [1-t/t_c]$ for $t \gg t^*$ and, then, the famous Family-Vicsek scaling \cite{FV} should be modified to
\begin{equation}
 w_2 = B t_c^{2\beta} f(t/t_c),
\end{equation}
where, according to Eq. \ref{eqanzatz}, $B = \Gamma^{2\beta}\langle \chi^2 \rangle_c$ and the scaling function should reads
\begin{equation}
f(x) \simeq \left \{
\begin{array}{ll}
x^{2 \beta}, & \text{for} \quad x \ll 0.8, \\
b (1-x), & \text{for} \quad 0.8 \ll x < 1,
\end{array}
\right.
\label{eqScaling2}
\end{equation}
with $b$ being an universal constant. Note that the last regime is hard to be observed, due to the stringent condition [$0.8 t_c \ll t < t_c$]. Moreover, as $t \rightarrow t_c$, the system sizes become so small that finite-size effects can hamper the scaling $w_2 \sim \langle L\rangle$. Indeed, only for the Etching model a reasonable evidence of this scaling behavior was found (see the insertion in Fig. \ref{fig1}a). Another possible explanation for the deviation in the scaling for the RSOS and SS models is the fact that, to respect the height difference restrictions in their interfaces, only some columns can be removed, making the remotions less random then in the Etching model. This becomes particularly relevant for small $L$ and may cause strong deviations in such regime.

Figure \ref{fig1}e presents the temporal variation of the skewness $S$ and kurtosis $K$ for the three models on ingrowing substrates ($\omega<0$). [Results (not shown) for the ratio $R$ display similar behaviors]. Clear plateaus are observed at the GOE values for $t \ll t^*$, which give place to a fast increasing behavior at long times $(t \gg t^*)$, similarly to those reported by FT \cite{Fukai2017}. Our main finding here, notwithstanding, is that $R$, $S$ and $K$, for a given model and $L_0$, have negligible differences up to $t \approx t^*$ for both $\omega>0$ and $\omega<0$ (as well as $\omega=0$), as shows Fig. \ref{fig1}e for the RSOS model. [Similar results are found for the other models]. This demonstrates that it does not matter if the size of the active growing zone of an 1D KPZ system is expanding or shrinking (or fixed), a transient GOE behavior will appear whenever the initial substrate size is large enough. As seem in Fig. \ref{fig1}f and clearly shown in Ref. \cite{Ismael14}, $S$ and $K$ for expanding systems ($\omega > 0$) start decreasing at $t \approx t_c$, because they converge to the GUE values when $t \rightarrow \infty$. Namely, by starting the regular outward growth ($\omega>0$) of a circular KPZ interface with $L_0 \gg 1$, a GOE-GUE crossover is observed \cite{Ismael14}. In the same way, for circular interfaces with inward growth ($\omega < 0$) a transient GOE regime is expected if $L_0 \gg 1$. However, in this case, a GOE-GUE crossover cannot be observed, because the interfaces become completely correlated before its (possible) onset.

\begin{figure}[t]
\includegraphics[width=8.5cm]{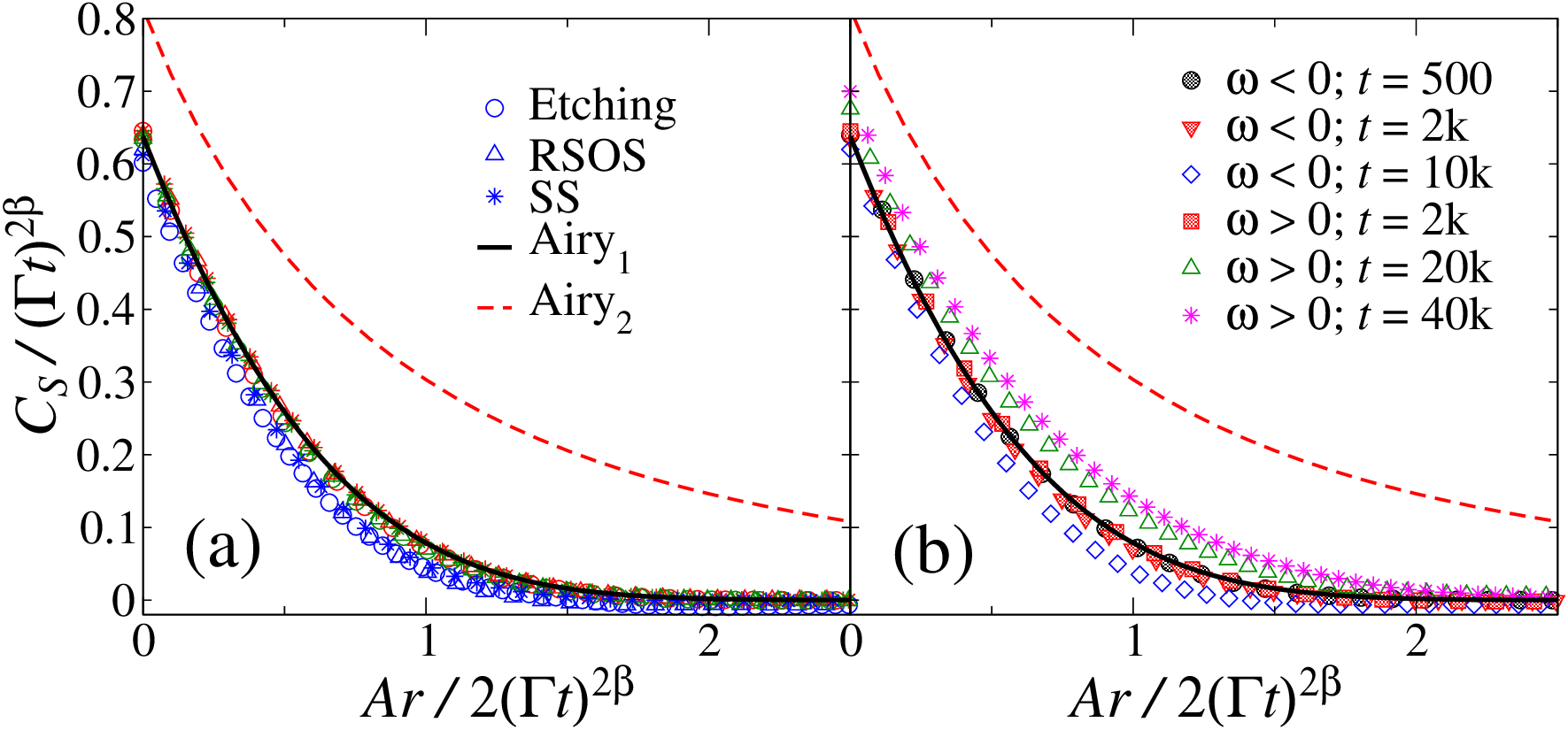}
\includegraphics[width=8.5cm]{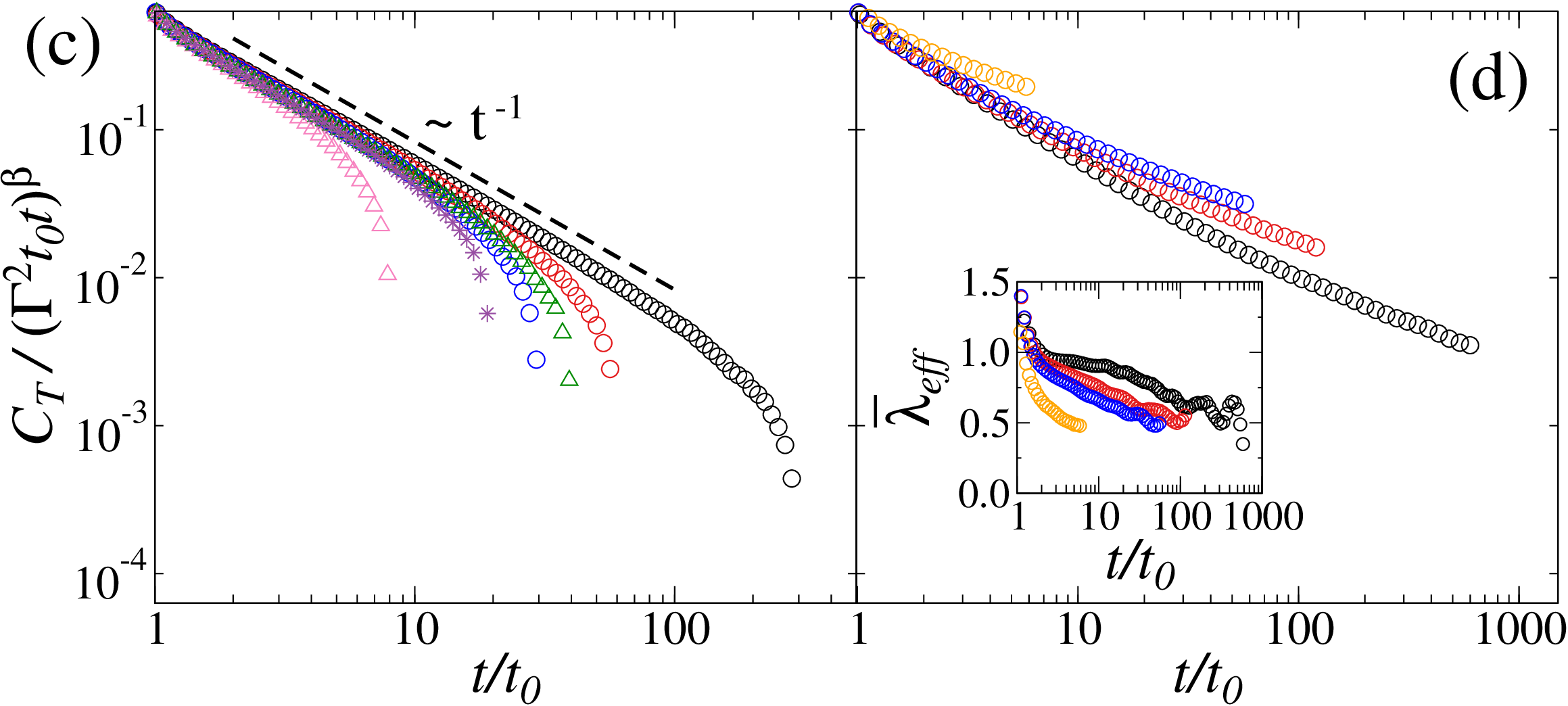}
\caption{(Color online) Rescaled spatial (a)-(b) and temporal (c)-(d) covariances. Data in (a) and (b) are for the same models/parameters in Figs. \ref{fig1}(e) and \ref{fig1}(f), respectively, for the times indicated by the arrows in those figures. Panel (c) shows temporal covariances for the RSOS (circles; for $\omega = -20$ and $L_0 =60\times10^4$), SS (triangles; for $\omega=-40$ and $L_0=16\times10^4$) and Etching (star; for $\omega=-40$ and $L_0=8\times10^4$) models, for $t_0 \in [100,1000]$. The data in (d) is for the RSOS model with $\omega=20$ and $L_0=16\times10^4$, for $t_0=100,500,1000,10000$. Effective exponents $\bar{\lambda}_{eff}$, calculated as the successive slopes of the curves in (d), are depicted in the insertion.}
\label{fig2}
\end{figure}

To confirm that the full flat statistics arises at the transient GOE regime, we calculate also the (2-point) spatial covariance $C_S(r,t) = \left\langle h(x,t) h(x+r,t) \right\rangle - \langle h \rangle^2 \simeq (\Gamma t)^{2 \beta} \Psi[A r^{2\alpha}/(\Gamma t)^{2 \beta}]$, which are displayed in Figs. \ref{fig2}a and \ref{fig2}b, for the models in Figs. \ref{fig1}e and \ref{fig1}f, respectively, for the growth times indicate by the arrows in these figures. Whenever $C_S$ is measured for times at the GOE plateaus, a nice collapse with the Airy$_1$ covariance is found, even for outgrowing systems ($\omega>0$). In this case, one can see the curves moving towards the Airy$_2$ covariance for long times ($t \gtrsim t_c$). On the other hand, for ingrowing systems ($\omega<0$) and $t \gtrsim t^*$, there is also a deviation from Airy$_1$, but in the opposite direction, as also found in \cite{Fukai2017}.

We analyze also the (2-point) temporal covariance $C_T(t,t_0) = \left\langle h(x,t_0) h(x,t) \right\rangle - \langle h \rangle^2 \simeq (\Gamma^2 t_0 t)^{\beta} \Phi(t/t_0)$, whose scaling function $\Phi(z)$ is expected to decay asymptotically as $\Phi(z) \sim z^{-\bar{\lambda}}$, with $\bar{\lambda}=\beta + 1/z$ for flat and $\bar{\lambda}=\beta$ for curved 1D interfaces \cite{Kallabis99,Singha2005}. In Figs. \ref{fig2}c and \ref{fig2}d examples of rescaled covariances for ingrowing and outgrowing systems are respectively shown, for large $L_0$. In both cases, at intermediate times the behavior is consistent with that expected for flat interfaces, namely, $\bar{\lambda} \approx 1$. For long times, however, the exponent approaches the value for curved 1D KPZ interfaces $\bar{\lambda} = 1/3$ for $\omega > 0$, as shows the effective exponents $\bar{\lambda}_{eff}$ displayed in the insertion of Fig. \ref{fig2}d. Actually, this flat-curved crossover manifests in two ways: {\it i}) For small $t_0$, a regime with $\bar{\lambda}\approx 1$ is observed, which is followed by a decreasing in $\bar{\lambda}_{eff}$ towards $\bar{\lambda} = 1/3$ as $t$ increases. {\it ii}) By increasing $t_0$ the initial regime $\bar{\lambda}\approx 1$ is lost and the large $t_0$ is the closer $\bar{\lambda}_{eff}$ is from $1/3$ at small ratios $t/t_0$. For $\omega<0$ a similar behavior is observed, however, rather than a crossover to the curved behavior, a fast decrease is found in the rescaled covariance for large $t/t_0$.

Altogether, the results above and those in Refs. \cite{Ismael14,Fukai2017} demonstrate that a full flat statistics shall appear in KPZ interfaces whenever $L_0 \gg 1$, regardless if they are growing inward or outward. In the last case, a crossover to the curved statistics occurs at $t\approx t_c$, which cannot be observed in the former case because the system correlates at $t^* < t_c$. Therefore, we are lead to inquiry how Yunker \textit{et al.} \cite{Yunker2013a} have found HDs consistent with GUE in the deposition of slight anisotropic colloidal particles at the edges of evaporating drops, once their system exhibits an inward growth. To understand this, we start remarking that, as stressed by some of us in Ref. \cite{tiago14Coloides}, all results in Ref. \cite{Yunker2013a} were obtained for the deposition of very few monolayers (MLs) of particles and, thus, they are probably far from any asymptotic regime (where GOE and GUE HDs live). Indeed, the apparent GUE behavior was observed for systems with average height $\langle h \rangle \lesssim 20\mu m$ in the deposition of particles with diameters $D \approx 1\mu m$. Namely, the deposits investigated in \cite{Yunker2013a} have less than 20 MLs and, in such regime, the HDs are expected to present only short-time transients and crossovers. 

\begin{figure}[t]
\includegraphics[width=8.5cm]{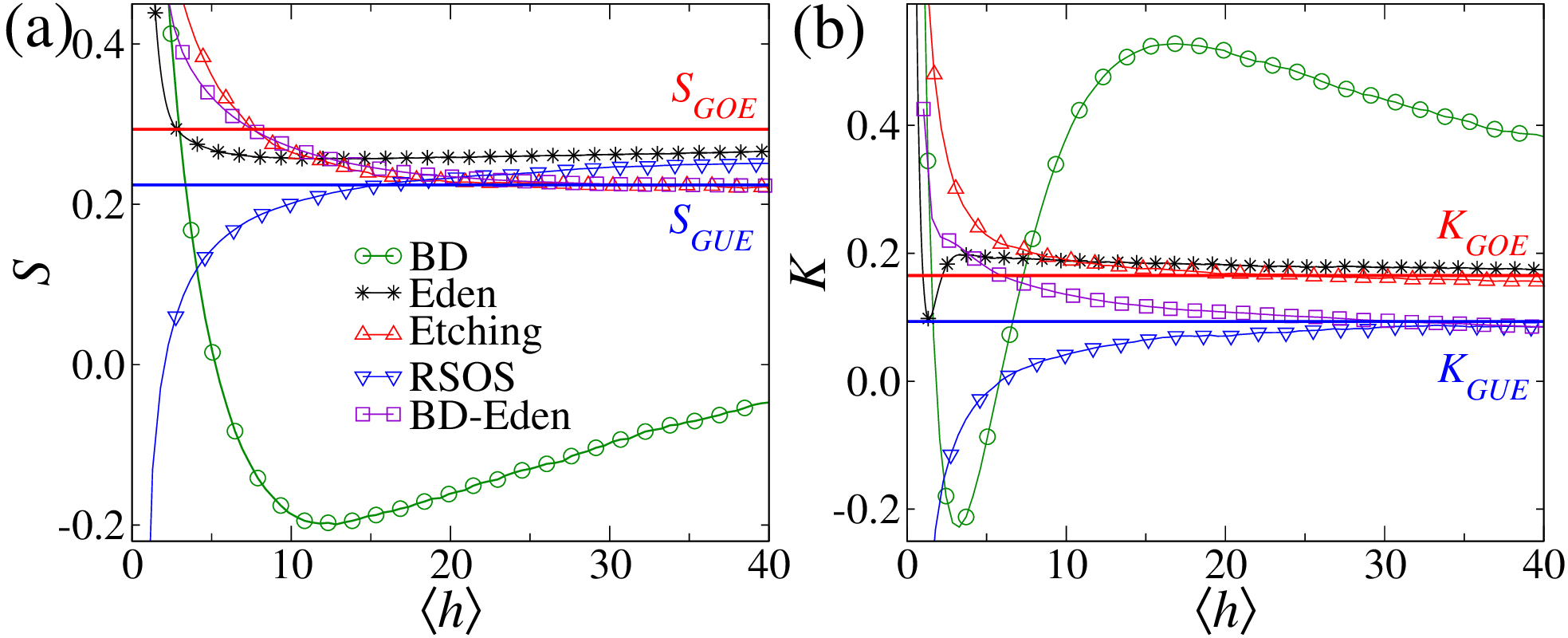}
\caption{(Color online) Short-time behavior of the HDs' cumulant ratios (a) skewness and (b) kurtosis as functions of the surface mean height $\langle h \rangle$, for several KPZ models, with $L_0=1600$, and $\omega=10$ [0] for BD, Etching and RSOS [Eden and BD-Eden] models. The horizontal solid lines indicate the values of these ratios for the GOE and GUE TW distributions.}
\label{fig3}
\end{figure}

To demonstrate this, we have performed simulations of three other KPZ models, focusing on their short-time HDs. In the ballistic deposition (BD) model, a particle is deposited at a (randomly chosen) site $i$ with $h_i \rightarrow max(h_{i-1},h_i+1,h_{i+1})$. In the Eden model \cite{Eden}, a deposit is created by adding particles at empty sites in its neighborhood. More specifically, we study here the so-called version B of the Eden model on the square lattice, starting from a line of particles at its bottom. The system evolves by randomly sorting occupied sites in the deposit which have empty NN sites and, then, placing a new particle in one of these empty sites at random. We investigate also a competitive BD-Eden model, where particles are deposited according to BD rule with probability $p$ or according to Eden rule with probability $(1-p)$, so that for $p=1$ ($p=0$) the simple BD (Eden) model is recovered. Figures \ref{fig3}a and \ref{fig3}b present, respectively, the skewness and kurtosis of the HDs for BD, Eden, Etching, RSOS and BD-Eden (with $p=0.35$) models, for deposition of very few MLs, comparable to those in the colloidal deposits. As expected, each model displays a different behavior and no trace of universality is found in this regime, indicating that we cannot draw reliable conclusions on the asymptotic HDs based on systems with $\sim 20$ MLs. Since each system has a different variation of $S$ and $K$ for small $\langle h \rangle$, there can exist even some ones that by chance agree with GUE during a short time interval. This is the case, for instance, in the BD-Eden model with probability $p=0.35$ (see Fig. \ref{fig3}), and the same thing can be happening in the colloidal deposition experiment. Investigation of this system for a larger number of MLs might confirm, or rule out this. We remark that, despite the approximated agreement with the experiments, we do not think that the BD-Eden model captures all key aggregation mechanisms from the colloidal system. Indeed, in the Supplemental video provided in \cite{Yunker2013a} one can see a much more complex aggregation behavior taking place there, with, e.g., the deposition of large clusters of particles, which can be also a relevant KPZ mechanism \cite{tiago07Graos,Ebrahiminejad}.

In summary, we have demonstrated that the flat statistics found by FT \cite{Fukai2017} in inward growth of liquid crystal experiments and Eden model simulations is a rather general crossover effect induced by the large initial system size $L_0$, which appears even in outward growth. Therefore, it is not a consequence of the inward growth or any effect of the sign of the initial curvature, as Ref. \cite{Fukai2017} may suggest. We remark that recent works on vapor deposition of CdTe films (which is known to be a 2D KPZ system \cite{Almeida14}) have reported crossovers from random to KPZ \cite{Almeida15} and from a \textit{pseudo}-steady state to KPZ (flat) statistics \cite{Almeida17}. Moreover, the crossover in the HDs from Edwards-Wilkinson \cite{EW} to KPZ class has been numerically investigated in some competitive growth models \cite{tiago13EWKPZ}. Thereby, the GOE-GUE crossover discussed here (for $\omega>0$) is only one example in a list of possible transient effects in KPZ systems. The GOE-correlated crossover observed here and by FT for $\omega<0$ is an expected result, for which a scaling relation for the interface width was introduced. Finally, a possible explanation for the apparent GUE HDs observed in the deposition of slight anisotropic colloidal particles was given, which possibly solves the controversy with the experimental results by FT, and points out that we have to be cautious when extracting information on asymptotic regimes from experimental interfaces.

\acknowledgments

We thank S. C. Ferreira for calling our attention for the inward growth problem, and the support from CNPq, Capes and FAPEMIG (brazilian agencies).

\bibliography{bibIngrowing}

\end{document}